\documentstyle[emulateapj,apjfonts,psfig]{article}

\begin{document}

\title{Constraining $\Omega_M$ and Dark Energy with Gamma-Ray Bursts}
\author{Z. G. Dai$^1$, E. W. Liang$^{1,2,3}$, and D. Xu$^1$}
\affil{$^1$Department of Astronomy, Nanjing University, Nanjing
210093, China; dzg@nju.edu.cn, ewliang@nju.eu.cn \\
$^2$Department of Physics, Guangxi University, Nanning 530004, China \\
$^3$National Astronomical Observatories/Yunnan Observatory, Chinese Academy of Sciences, Kunming 650011, China \\
{\em To appear in the Astrophysical Journal Letters} }

\begin{abstract}
An $E_{\gamma,{\rm jet}}\propto {E'_p}^{1.5}$ relationship with a small scatter for current $\gamma$-ray burst (GRB)
data was recently reported, where $E_{\gamma,{\rm jet}}$ is the beaming-corrected $\gamma$-ray energy and $E'_p$ is the
$\nu F_\nu$ peak energy in the local observer frame. By considering this relationship for a sample of 12 GRBs with
known redshift, peak energy, and break time of afterglow light curves, we constrain the mass density of the universe
and the nature of dark energy. We find that the mass density $\Omega_M=0.35\pm^{0.15}_{0.15}$ (at the $1\sigma$
confident level) for a flat universe with a cosmological constant, and the $w$ parameter of an assumed static
dark-energy equation of state $w=-0.84\pm^{0.57}_{0.83}$ ($1\sigma$). Our results are consistent with those from type
Ia supernovae. A larger sample established by the upcoming {\em Swift} satellite is expected to provide further
constraints.
\end{abstract}


\section {Introduction}

Type Ia supernovae (SNe Ia) have been playing an important role in modern cosmology. Early observations of SNe Ia at
redshift $z<1$ strongly suggest that the expansion of the universe at the present time is accelerating (Riess et al.
1998; Perlmutter et al. 1999). Since then, the nature of dark energy (with negative pressure) that drives cosmic
acceleration has been one of the greatest mysteries in modern cosmology (for reviews see Peebles \& Ratra 2003;
Padmanabhan 2003). Recent observations of 16 higher-redshift (up to $z\simeq 1.7$) SNe Ia present conclusive evidence
that the universe had once been decelerating (Riess et al. 2004). These newly-discovered objects, together with
previous reported SNe Ia, have been used to provide further constraints on both the expansion history of the universe
and the equation of state (EOS) of a dark energy component (Riess et al. 2004; Wang \& Tegmark 2004; Daly \& Djorgovski
2004; Feng, Wang \& Zhang 2004).

$\gamma$-ray bursts (GRBs) are the brightest electromagnetic explosions in the universe. It has been widely believed
that they should be detectable out to very high redshifts (Lamb \& Reichart 2000; Ciardi \& Loeb 2000; Bromm \& Loeb
2002). $\gamma$-ray photons with energy from tens of keV to MeV, if produced at high redshifts, suffer from no
extinction before they are detected. These advantages would make GRBs an attractive probe of the universe. Schaefer
(2003) derived the luminosity distances of 9 GRBs with known redshifts by using two luminosity indicators (the spectral
lag and the variability). He obtained the first GRB Hubble diagram with the $1\sigma$ constraint on the mass density
$\Omega_M<0.35$.

A correlation between the isotropic-equivalent $\gamma$-ray energy ($E_{\gamma,{\rm iso}}$) and the $\nu F_\nu$ peak
energy ($E'_p$) in the local observer frame, $E'_p\propto E_{\gamma,{\rm iso}}^{1/2}$, was discovered by BeppoSAX
observations (Amati et al. 2002; Yonetoku et al. 2004), and confirmed by HETE-2 observations (Sakamoto et al. 2004;
Lamb et al. 2004). It not only holds among BATSE GRBs (Lloyd-Ronning \& Ramirez-Ruiz 2002) but also within one GRB
(Liang, Dai \& Wu 2004). Its possible explanations include the synchrotron mechanism in relativistic shocks (Zhang \&
M\'esz\'aros 2002; Dai \& Lu 2002) and the emission from off-axis relativistic jets (Yamazaki, Ioka \& Nakamura 2004;
Eichler \& Levinson 2004). However, the dispersion around this correlation is too large to obtain useful information on
the universe from the current GRB sample.

Ghirlanda, Ghisellini \& Lazzati (2004) recently found a new relationship between the beaming-corrected $\gamma$-ray
energy ($E_{\gamma,{\rm jet}}$) and the local-observer peak energy, $E_{\gamma,{\rm jet}}\propto {E'_p}^{1.5}$, with a
small scatter for current GRB data, suggesting that GRBs are a promising probe of the universe. In principle, this
relationship can be derived from the $E'_p\propto E_{\gamma,{\rm iso}}^{1/2}$ correlation combined with the afterglow
jet model. In this Letter, we constrain the mass density of the universe and the nature of dark energy by considering
this relationship with a sample of 12 GRBs with known redshift, peak energy, and break time of afterglow light curves.
We show that GRBs appear to provide an independent and interesting probe of fundamental quantities of the universe.

\section{Sample Selection and Standard Candles}

By searching for GRBs in the literature, we have found total 14 bursts of which redshift $z$, observed peak energy
$E_p$, and break time $t_{\rm j}$ of afterglow light curves are available. Table 1 lists a sample of 12 GRBs, but the
other two events, GRBs 990510 and 030226, are not included. The reason is as follows: the analysis of this Letter and
Ghirlanda et al. (2004) is based on the afterglow jet model (Rhoads 1999; Sari, Piran \& Halpern 1999). In this model,
a relativistic jet, after emitting a fraction $\eta_\gamma$ of its kinetic energy into prompt $\gamma$ rays, expands in
a homogeneous medium with number density of $n$. As the jet sweeps up more and more medium matter, its Lorentz factor
declines. When the Lorentz factor equals the inverse of the jet's half opening angle $\theta$, the afterglow light
curve presents a break. However, this model cannot well fit the afterglow data of these two bursts, because the
predicted break spans about two orders of magnitude in time when light travel time effects are taken into account, and
thus the theoretical light curve is too smooth to be consistent with the observed sharpness (Rhoads \& Fruchter 2001;
Wei \& Lu 2002). For GRB 990510, the required spectral index of the electrons is less than 2, being inconsistent with
the shock acceleration theory (Wei \& Lu 2002). In addition, the afterglow data of GRB 030226 suggest that its
environment might be a low-density wind rather than a constant-density medium, also conflicting with the model (Dai \&
Wu 2003).

According to the afterglow jet model (Sari et al. 1999), the jet's half opening angle is given by $\theta =
0.161(1+z)^{-3/8}t_{{\rm j},d}^{3/8}E_{\gamma,{\rm iso},52}^{-1/8}n_0^{1/8}\eta_\gamma^{1/8}$, where $E_{\gamma,{\rm
iso},52}=E_{\gamma,{\rm iso}}/10^{52}{\rm ergs}$, $t_{{\rm j},d}=t_{\rm j}/1\,{\rm day}$, $n_0=n/1\,{\rm cm}^{-3}$, and
$\eta_\gamma=0.2$ (Frail et al. 2001). Only for few bursts in Table 1 the medium density was obtained from broadband
modelling of the afterglow emission (e.g., Panaitescu \& Kumar 2002). For those bursts with unknown $n$, we assume the
median density $n\simeq 3\,{\rm cm}^{-3}$ as in Ghirlanda et al. (2004). The isotropic-equivalent $\gamma$-ray energy
of a GRB is calculated by
\begin{equation}
E_{\gamma,{\rm iso}}=\frac{4\pi d_L^2S_\gamma k}{1+z},
\end{equation}
where $S_\gamma$ is the fluence (in units of erg\,cm$^{-2}$) received in some observed bandpass and $k$ is the factor
that corrects the observed fluence to the standard rest-frame bandpass (1-$10^4$ keV) (Bloom, Frail \& Sari 2001). For
a Friedmann-Robertson-Walker (FRW) cosmology with mass density $\Omega_M$ and vacuum energy density $\Omega_\Lambda$,
the luminosity distance in equation (1) is
\begin{eqnarray}
d_L & = & c(1+z)H_0^{-1}|\Omega_k|^{-1/2}{\rm sinn}\{|\Omega_k|^{1/2}\nonumber \\ & & \times
\int_0^zdz[(1+z)^2(1+\Omega_Mz)-z(2+z)\Omega_\Lambda]^{-1/2}\},
\end{eqnarray}
where $c$ is the speed of light and $H_0\equiv 100h\,\,{\rm km}\,{\rm s}^{-1}\,{\rm Mpc}^{-1}$ is the present Hubble
constant (Carroll, Press \& Turner 1992). In equation (2), $\Omega_k=1-\Omega_M-\Omega_\Lambda$, and ``sinn" is sinh
for $\Omega_k>0$ and sin for $\Omega_k<0$. For $\Omega_k=0$, equation (2) turns out to be $c(1+z)H_0^{-1}$ times the
integral. In this section, we assume a flat universe (i.e., $\Omega_k=0$) because of both an expected consequence of
inflation and the observed characteristic angular size scale of the cosmic microwave background fluctuations (Spergel
et al. 2003 and references therein).

From equations (1) and (2), we obtain the beaming-corrected $\gamma$-ray energy $E_{\gamma,{\rm jet}}=(1-\cos
\theta)E_{\gamma,{\rm iso}}$, that is,
\begin{equation}
E_{\gamma,{\rm jet}}\simeq 1.30\times 10^{50}(1+z)^{-3/4}t_{{\rm j},d}^{3/4}E_{\gamma,{\rm
iso},52}^{3/4}n_0^{1/4}\eta_\gamma^{1/4}\,{\rm ergs}.
\end{equation}
Figure 1 plots $E_{\gamma,{\rm jet}}$ versus $E'_p$ for the GRB sample listed in Table 1, with $\Omega_M=0.27$,
$\Omega_\Lambda=0.73$ and $h=0.71$. We find that $E_{\gamma,{\rm jet}}$ and $E'_p$ are strongly correlated with a
correlation coefficient $r_s=0.99\pm 0.08$ (with a probability of $<10^{-4}$). The best fit is $(E_{\gamma,{\rm
jet}}/10^{50}{\rm ergs})=(1.12\pm0.12)(E'_p/100\,{\rm keV})^{1.50\pm 0.08}$ with a reduced $\chi^2_{\rm dof}=0.53$. We
note this power to be insensitive to $\Omega_M$. In addition, although the peak energy $E'_p$ and the low-energy
spectral index $\alpha$ in Table 1 appear to evolve with redshift (Amati et al. 2002), this evolution doesn't affect
the above relation as shown in Figure 1. These results imply that GRBs are standard candles.

\section{Hubble Diagram and Cosmological Constraints}

We first derive the observed luminosity distance from the GRB sample. Considering a relationship $(E_{\gamma, {\rm
jet}}/10^{50}{\rm ergs})=C(E'_p/100\,{\rm keV})^{1.5}$ (where $C$ is a dimensionless parameter), we obtain
\begin{equation}
d_L=2.37\times 10^{23}\frac{(1+z)^2C^{2/3}E_p}{(kS_\gamma t_{{\rm j},d})^{1/2}(n_0\eta_\gamma)^{1/6}}\,{\rm cm},
\end{equation}
where $E_p=E'_p/(1+z)$ is in units of keV. Thus, the observed distance modulus of a GRB is $\mu_{\rm ob}=5\log
(d_L/10\,{\rm pc})$ with an error of
\begin{equation}
\sigma_{\mu_{\rm ob}} = 2.17 \left[
      \left(\frac{\sigma_{E_{p}}}{E_{p}}\right)^2+
      \left(\frac{\sigma_{S_{\gamma}}}{2S_{\gamma}}\right)^2 +
      \left(\frac{\sigma_{t_{{\rm j}}}}{2t_{{\rm j}}}\right)^2 +
      \left(\frac{\sigma_{n}}{6n}\right)^2 \right]^{1/2},
\end{equation}
where $\sigma_{E_{p}}$, $\sigma_{S_{\gamma}}$, $\sigma_{t_{{\rm j}}}$, and $\sigma_{n}$ are the errors in the peak
energy, fluence, break time and medium density of the GRB, respectively.

We plot a Hubble diagram of our GRB sample in Figure 2 based on equations (4) and (5). This figure also presents a
Hubble diagram of the current SNe Ia sample. Both Hubble diagrams are consistent with each other. However, GRBs and SNe
Ia have mean uncertainties of $0.09$ and $0.05$ in the log of the distance, respectively, and thus GRBs are about twice
worse in accuracy than SNe.

For an FRW cosmology with $\Omega_M$ and $\Omega_\Lambda$, equation (2) gives the theoretical distance modulus
$\mu_{\rm th}=5\log (d_L/10\,{\rm pc})$. The likelihood for these cosmological parameters can be determined from a
$\chi^2$ statistic, where
\begin{equation}
\chi^2(h,\Omega_M,\Omega_\Lambda;C)=\sum_i\frac{[\mu_{\rm th}(z_i;h,\Omega_M,\Omega_\Lambda)-\mu_{{\rm
ob},i}(C)]^2}{\sigma_{\mu_{{\rm ob},i}}^2}.
\end{equation}
We consider all possible values of the parameters $h$ and $C$ to be $h\in (0.68, 0.75)$ (Bennett et al. 2003) and
$C\in(1.00, 1.24)$ (see \S 2). The confidence regions in the $\Omega_M$-$\Omega_\Lambda$ plane can be found through
marginalizing the likelihood functions over $h$ and $C$ (i.e., integrating the probability density $P\propto
e^{-\chi^2/2}$ for all values of $h$ and $C$). We plot contours of likelihood (from $1\sigma$ to $3\sigma$) for unknown
curvature $\Omega_k$ in Figure 3. As shown for a flat universe, with the current sample, $\Omega_M<0.62$ (at the
$2\sigma$ confidence level), and the $1\sigma$ contour contains the $(\Omega_M, \Omega_\Lambda)=(0.27, 0.73)$ point
corresponding to the ``concordance" model. We measure $\Omega_M=0.35\pm^{0.15}_{0.15}$ ($1\sigma$).

There are several alternative approaches to calculate the luminosity distance (also see Riess et al. 2004). We here
consider a flat universe and a constant EOS, $w=P_{\rm DE}/\rho_{\rm DE}c^2$, of a dark energy component (Garnavich et
al. 1998; Perlmutter et al. 1999). In this case we have
\begin{eqnarray}
d_L & = & c(1+z)H_0^{-1}\int_0^zdz[\Omega_M(1+z)^3\nonumber \\
& & +(1-\Omega_M)(1+z)^{3(1+w)}]^{-1/2}.
\end{eqnarray}
Figure 4 presents contours of likelihood in the $\Omega_M$-$w$ plane (after marginalizing over $h$ and $C$). The solid
contours consider a prior of $\Omega_M=0.27\pm 0.04$ by assuming its Gaussian distribution, similar to Riess et al.
(2004). We see $w=-0.84\pm^{0.57}_{0.83}$ ($1\sigma$), which is consistent with the value of $w$ expected for a
cosmological constant (i.e., $w=-1$).

\section{Conclusions}

The $E_{\gamma,{\rm jet}}\propto {E'_p}^{1.5}$ relationship with a small dispersion was reported by Ghirlanda et al.
(2004) and confirmed in this Letter. The advantages of considering this relationship as a probe of the universe are (1)
that GRBs have been detected at redshifts up to $z\simeq 4.5$, (2) that $\gamma$ rays suffer from no extinction, and
(3) that we don't think of luminosity evolution. These advantages led us to constrain the mass density of the universe
and the nature of dark energy. We found that the mass density $\Omega_M=0.35\pm^{0.15}_{0.15}$ ($1\sigma$) for a flat
universe with a cosmological constant, and the $w$ parameter of the dark-energy EOS $w=-0.84\pm^{0.57}_{0.83}$
($1\sigma$). Riess et al. (2004) measured $\Omega_M=0.29\pm^{0.05}_{0.03}$ and $w=-1.02\pm ^{0.13}_{0.19}$ ($1\sigma$)
for the current SNe Ia sample. Therefore, our results are consistent with those from SNe Ia.

The upcoming {\em Swift} satellite with an energy range of $0.2-150$ keV will be scheduled for launch in 2004 September
(Gehrels et al. 2004). {\em Swift} is expected (1) to detect more than 100 bursts per year, (2) to observe X-ray and
UV/optical afterglows at times of 1 minute to several days after the burst, and (3) to detect very-high-redshift GRBs.
Thus, it is expected that a larger sample of GRBs established by {\em Swift} provides a further probe of the universe.
Such a probe opens up a new window on the cosmic distance scale far beyond the reach of SNe Ia. We call this research
field {\em GRB cosmology}, corresponding to the well-known supernova cosmology.

\acknowledgments We thank the referee for constructive suggestions, and Bing Zhang and Xinmin Zhang for valuable
comments. This work was supported by the National Natural Science Foundation of China (grants 10233010 and 10221001),
the Ministry of Science and Technology of China (NKBRSF G19990754), the Natural Science Foundation of Yunnan
(2001A0025Q), and the Research Foundation of Guangxi University.

\clearpage

\begin{table}
\caption{Sample of $\gamma$-ray bursts}
\begin{tabular}{cccccccccc}
\hline\hline GRB & redshift & $E_p(\sigma_{E_p})\tablenotemark{a}$ & $\alpha\tablenotemark{a}$ &
$\beta\tablenotemark{a}$ & $S_\gamma(\sigma_{S_\gamma}) \tablenotemark{b}$ & range\tablenotemark{b} & $t_{\rm
j}(\sigma_{t_{\rm j}})\tablenotemark{c}$&$n (\sigma
_n)\tablenotemark{d}$ & ref\tablenotemark{e}\\ & & (keV) & & & (erg\,cm$^{-2})$ & (keV) & (days) & (cm$^{-3})$ & \\
\hline
970828...... &0.957&297.9(59.3)&-0.7&-2.07&9.6E-5(0.9)&20-2000&2.2(0.4)&3(0.33)&1,13,19\\
980703...... &0.966&255.3(50.9)&-1.31&-2.39&2.3E-5(0.2)&20-2000&3.4(0.5)&28(10)&2,13,20,20\\
990123...... &1.6&780.8(61.9)&-0.89&-2.45&3.0E-4(0.4)&40-700&2.04(0.46)&3(0.33)&3,14,21\\
990705...... &0.843&188.8(15.2)&-1.05&-2.2&7.5E-5(0.8)&40-700&1.0(0.2)&3(0.33)&4,14,22\\
990712...... &0.43&65.0(10.5)&-1.88&-2.48&6.5E-6(0.3)&40-700&1.6(0.2)&3(0.33)&5,14,23\\
991216...... &1.02&317.3(63.4)&-1.234&-2.18&1.9E-4(0.2)&20-2000&1.2(0.4)&4.7(2.3)&6,13,24,25\\
011211...... &2.14&59.2(7.6)&-0.84&-2.3&5.0E-6(0.5)&30-400&1.50(0.02)&3(0.33)&7,14,26\\
020124...... &3.2&110.0(22.0)&-1&-2.3&6.8E-6(0.7)&30-400&3.0(0.4)&3(0.33)&8,15,27\\
020405...... &0.69&192.5(53.8)& 0& -1.87&7.4E-5(0.7)&15-2000&1.67(0.52)&3(0.33)&9,16,28\\
020813...... &1.25&211.0(42.0)&-1.05&-2.3&1.0E-4(0.1)&30-400&0.43(0.06)&3(0.33)&10,17,29\\
030328...... &1.52&109.9(21.8)&-1&-2.3&2.6E-5(0.2)&30-400&0.8(0.1)&3(0.33)&11,17,27\\
030329...... &0.1685&67.6(2.6)&-1.26&-2.28&1.1E-4(0.1)&30-400&0.5(0.1)&1(0.11)&12,18,30\\
\hline
\end{tabular}
\tablecomments{(a) The spectral parameters fitted by the Band function; (b) the fluence and error observed in the
corresponding energy range; (c) the observed break time and error of the afterglow light curve; (d) the medium density
and error from afterglow fittings, if no available the value of $n$ taken to be $3\pm0.33$ cm $^{-3}$; (e) references
in order for redshift, spectral data, $t_{\rm j}$, and $n$.}

\tablerefs{(1) Djorgovski et al. 1998a; (2) Djorgovski et al. 1998b; (3) Hjorth 1999; (4) Amati et al. 2000; (5) Galama
et al. 1999; (6) Vreeswijk et al. 1999a; (7) Andersen et al. 2000; (8) Hjorth et al. 2003; (9) Masetti et al. 2002;
(10) Price et al. 2002; (11) Rol et al. 2003; (12) Greiner et al. 2003; (13) Jimenez et al. 2001; (14) Amati et al.
2002; (15) Barraud et al. 2003; (16) Price et al. 2003; (17) Atteia 2003; (18) Vanderspek et al. 2004; (19) Djorgovski
et al. 2001; (20) Frail et al. 2003; (21) Kulkarni et al. 1999; (22) Masetti et al. 2000; (23) Bjornsson et al. 2001;
(24) Halpern et al. 2000; (25) Panaitescu \& Kumar 2002; (26) Jakobsson et al. 2003; (27) Ghirlanda et al. 2004; (28)
Price et al. 2003; (29) Barth et al. 2003; (30) Berger et al. 2004}
\end{table}
\clearpage

\begin{figure}
\plotone{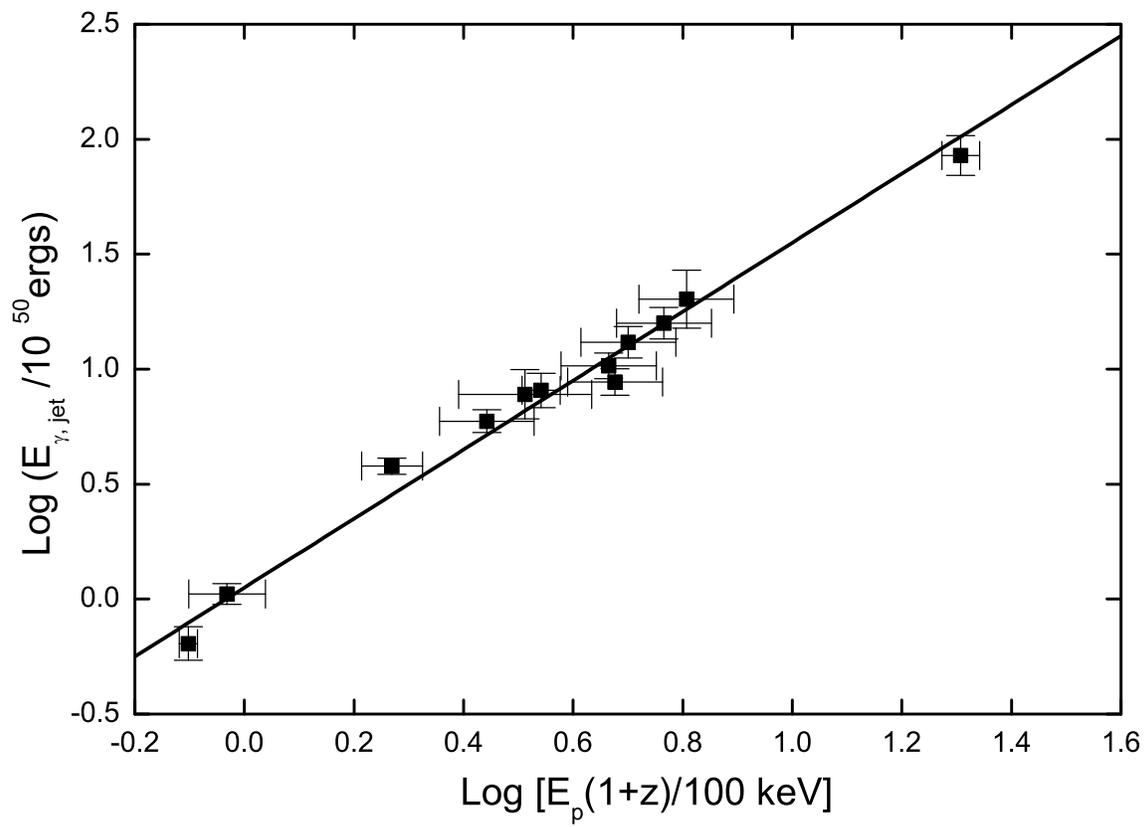} \caption{Beaming-corrected $\gamma$-ray energy versus local-observer peak energy for the GRB sample
listed in Table 1. The line is the best fit. \label{fig1}}
\end{figure}

\begin{figure}
\plotone{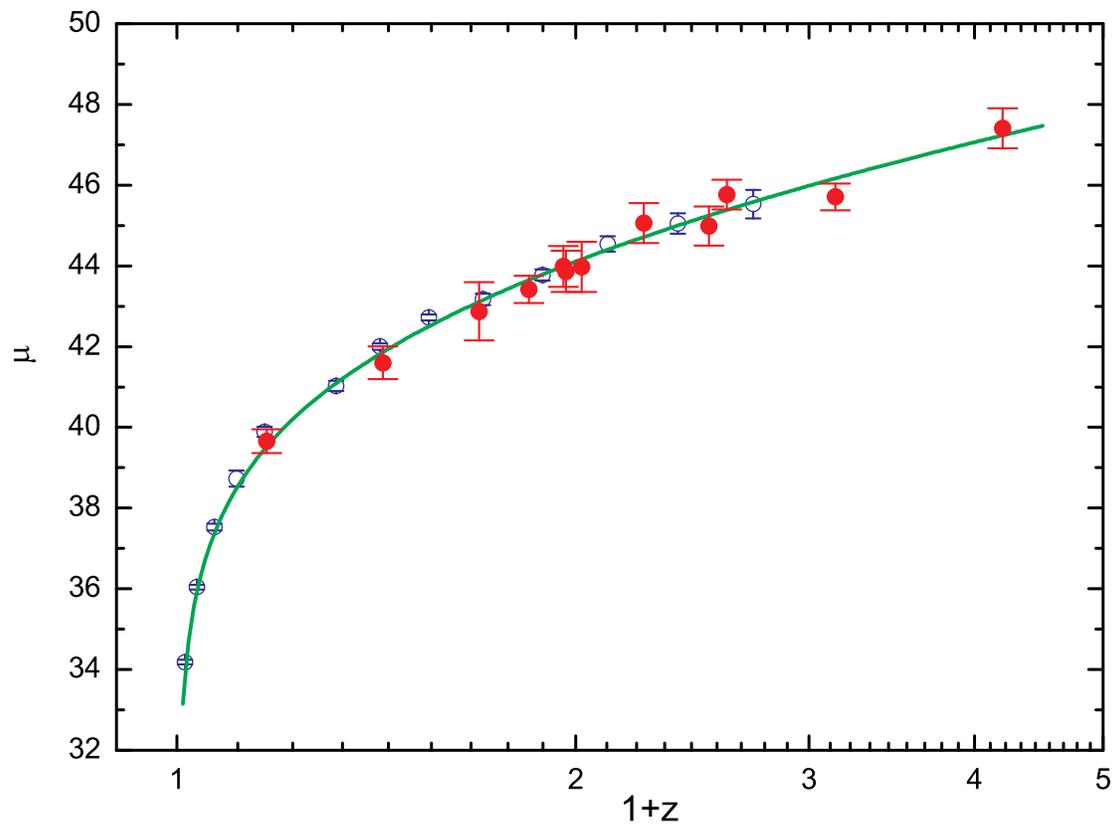} \caption{Hubble diagrams for the GRB sample ({\em filled} symbols, for $C=1.12$) and the binned SN
Ia data ({\em open} symbols) from Riess et al. (2004). The line corresponds to a flat cosmology with $\Omega_M=0.27$
and $h=0.71$. \label{fig2}}
\end{figure}

\begin{figure}
\plotone{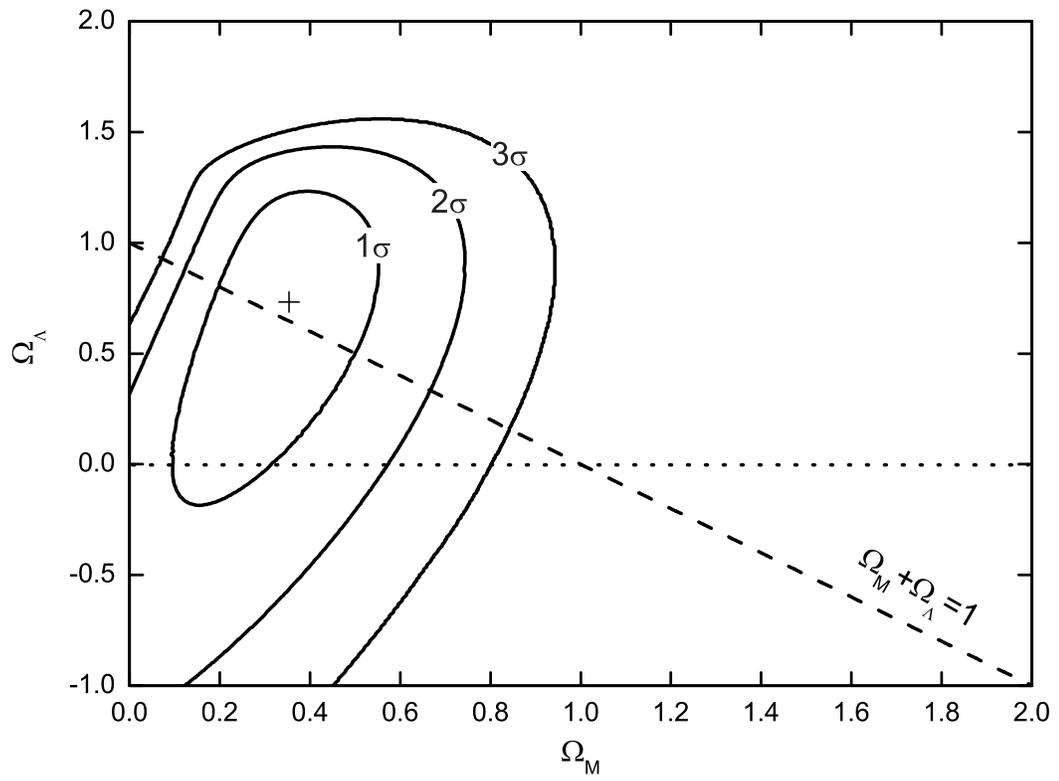} \caption{Contours of likelihood in the $\Omega_M$-$\Omega_\Lambda$ plane. The cross indicates the
best fit, and the dashed line is for a flat universe. \label{fig3}}
\end{figure}

\begin{figure}
\plotone{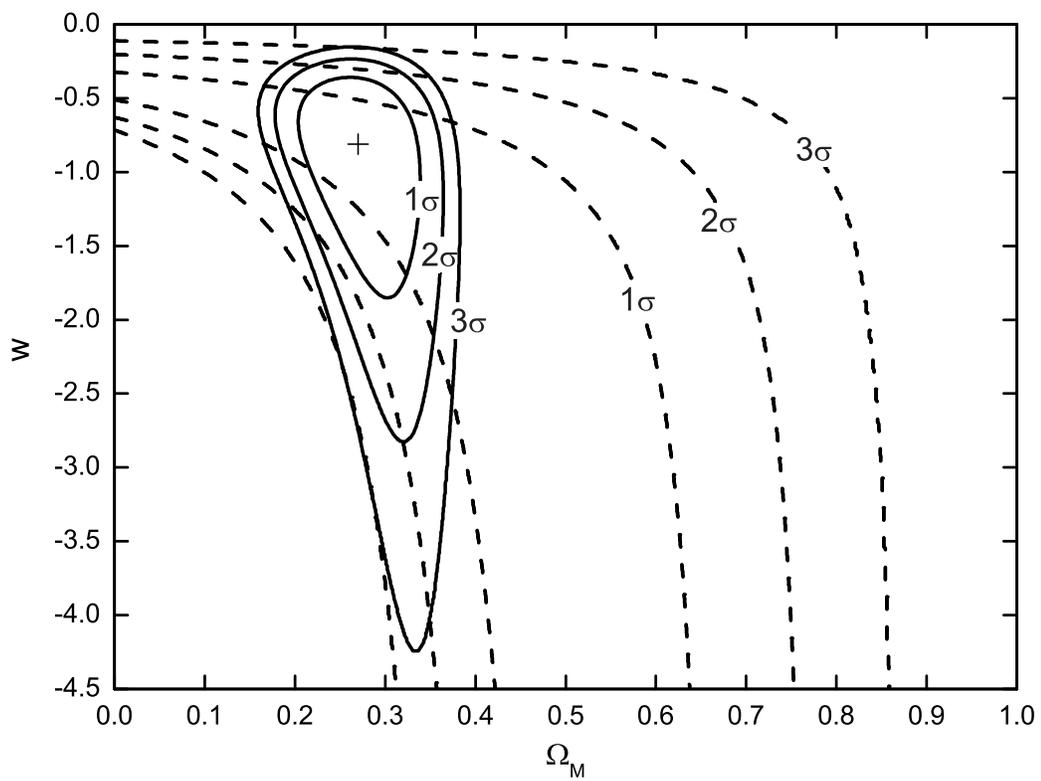} \caption{Contours of likelihood from the GRB sample (dashed lines) in the $\Omega_M$-$w$ plane. The
solid contours consider a prior of $\Omega_M=0.27\pm 0.04$. The cross indicates the best fit. \label{fig4}}
\end{figure}

\end{document}